\documentclass{article}
\usepackage{spconf,amsmath,graphicx}
\usepackage{amssymb}
\DeclareMathOperator*{\argmin}{arg\,min}

\DeclareMathOperator*{\tr}{tr}
\DeclareMathOperator*{\E}{E}
\DeclareMathOperator*{\sign}{\textrm{sign}}
\DeclareMathOperator*{\Q}{\mathcal{Q}}

\DeclareMathOperator*{\diag}{\textrm{diag}}

\newcommand{\vect}[1]{\mathbf{#1}}
\newcommand{\matt}[1]{\mathbf{#1}}
\newcommand{\jim}{\mathrm{j}\,}
\newcommand{\T}{\operatorname{\mathrm{T}}}

\title{MMSE Precoder for Massive MIMO using 1-bit Quantization}
\name{Ovais Bin Usman, Hela Jedda, Amine Mezghani and Josef A. Nossek}% \thanks{Thanks to XYZ agency for funding.}}
\address{Institute for Circuit Theory and Signal Processing\\
		Technische Universit\"at M\"unchen, 80290 Munich, Germany\\
		Email:\{hela.jedda, josef.a.nossek\}@tum.de}

\begin{document}
%\ninept
%
\maketitle
\begin{abstract}
We propose a novel linear minimum-mean-squared-error (MMSE) precoder design for a downlink (DL) massive multiple-input-multiple-output (MIMO) scenario. For economical and computational efficiency reasons low resolution 1-bit digital-to-analog (DAC) and analog-to-digital (ADC) converters are used. This comes at the cost of performance gain that can be recovered by the large number of antennas deployed at the base station (BS) and an appropiate precoder design to mitigate the distortions due to the coarse quantization. The proposed precoder takes the quantization non-linearities into account and is split into a digital precoder and an analog precoder. We formulate the two-stage precoding problem such that the MSE of the users is minimized under the 1-bit constraint. In the simulations, we compare the new optimized precoding scheme with previously proposed linear precoders in terms of uncoded bit error ratio (BER).
\end{abstract}
\begin{keywords}
Massive MIMO, Precoding, 1-bit quantization, Transmit signal processing
\end{keywords}
\section{Introduction}
\label{sec:intro}

The massive MIMO system, or named large-scale antenna system has been seen as a promising technology for the next generation wireless communication systems \cite{Marzetta2010, Bjornson2013}. The huge increase in the number of antennas at BS can improve spectral efficiency (SE), energy efficiency (EE) and reliability.
The BS with large number of antennas, say 100 antennas or more, simultaneously serves
a much smaller number of single-antenna users. With the knowledge of CSI at the BS (CSIT), this large spatial DoF of massive MIMO systems can be exploited to significantly increase the spatial multiplexing/diversity gain using
MU-MIMO precoding \cite{Peel2005, Gershman2010}. The linear precoders, such as MF, ZF \cite{Rusek2013} and the regularized zero-forcing (RZF) scheme \cite{Peel2005} are shown to be near-optimal. Thus, it is more practical to use low-complexity linear precoding techniques in massive MIMO systems. Therefore, we mainly focus on linear
precoding techniques in this work. The price to pay for massive MIMO systems is increased complexity of the hardware (number of radio frequency (RF) and ADC/DAC chains) and the signal processing and resulting increased energy consumption at the transmitter \cite{Rusek2013}. Several approaches are considered in the literature 
to decrease the power consumption such as spatial modulation \cite{Renzo2014}, load modulation \cite{MullerSedaghat}, the use of parasitic antennas \cite{Kalis2012} and the use of low-cost transceivers \cite{Bjornson2015}. 
%The authors in \cite{AnLiu2014} proposed a two-stage precoding scheme where the number of RF chains is much less than the number of the BS antennas. The users are partitioned into non-overlapping clusters. The precoder is split into a baseband precoder with a much less number of RF chains for intra-cluster spatial multiplexing and an RF precoder using the RF phase shifting network for inter-cluster interference mitigation.

One attractive solution  to overcome the issues of high complexity and high energy consumption associated with massive MIMO, 
is the use of very low resolution ADCs and DACs. The power consumption of the ADC and the DAC, one of the most power-hungry devices, can be reduced exponentially by decreasing the resolution \cite{Svensson2006} and 1-bit quantization can
drastically simplify other RF-components, e.g., amplifiers and mixers. Therefore, we design an MMSE linear precoder in a DL massive MIMO scenario where the resolution of the DACs and ADCs is restricted to 1 bit. This precoder design aims at mitigating the distortions due to the coarse quantization in addition to inter-user interference (IUI). A similar work has been presented in \cite{Mezghani2009}, where the authors optimize first the quantizer's levels and then give a closed-form expression of an MMSE precoder that takes into account the quantizer non-linearities. However, only quantization at the transmitter was considered. In this contribution, we do not optimize the quantizer. The quantizer in our work has constant levels. But we introduce a second precoding stage in the analog domain after the quantizer to minimize the distortions due to 1-bit DAC/ADC in i.i.d. complex Gaussian channels. The proposed two-stage precoder is designed based on iterative methods. We assume perfect CSIT and study how the new precoder scheme is improving the BER compared to the precoder introduced in \cite{Mezghani2009}.

This paper is organized as follows: in Section \ref{sec:sysmodel} the system model is presented. In Section \ref{sec:onebitQ} some derivations related to the 1-bit quantization are introduced. In Section \ref{sec:optproblem} we formulate our optimization problem and show the derivations and the corresponding solution. In Sections \ref{sec:simresults} and \ref{sec:concl} we interpret the simulation results and summarize this work.

Notation: Bold letters indicate vectors and matrices, non-bold letters express scalars. The operators $(.)^{*}$, $(.)^{\rm T}$, $(.)^{\rm H}$ and $\E\left[.\right]$ stand for complex conjugation, the transposition, Hermitian transposition and the expectation, respectively. The $n \times n$ identity matrix is denoted by $\mathbf{I}_{n}$ while the zeros (ones) matrix with $n$ rows and $m$ columns is defined as $\mathbf{0}_{n,m}$ ($\mathbf{1}_{n,m}$). We define $\left(\bullet\right)_{R} = \Re \lbrace \bullet \rbrace$, $\left(\bullet\right)_{I} = \Im \lbrace \bullet \rbrace$ and $\Q(x) = \sign(x_{R}) + \jim \sign(x_{I})$. Additionally, $\diag(\matt{A})$ denotes a diagonal matrix containing only the diagonal
elements of $\matt{A}$. $\sigma_{\alpha}$ and $\rho_{\alpha \beta}$ denote the standard deviation of $\alpha$ and the correlation coefficient between $\alpha$ and $\beta$, respectively. For a circular distributed Gaussian complex-valued signal $\alpha$ we have $\sigma_{{\alpha}_R} = \sigma_{{\alpha}_I}$.

\section{System Model}
\label{sec:sysmodel}
\begin{figure}
  \centering
  \centerline{\includegraphics[width=8.5cm]{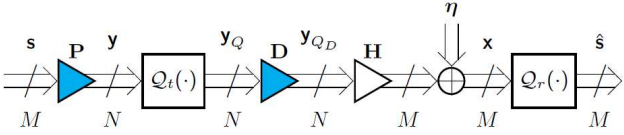}}
  \caption{System Model}\medskip
  \label{fig:sysmodel}
\end{figure}

We consider a massive MIMO downlink scenario as depicted in Fig. \ref{fig:sysmodel}. The BS with $N$ antennas serves $M$ single-antenna users, where $N \gg M$. The signal vector $\vect{s} \in \mathcal{O}^{M}$ contains the data symbols for each of the $M$ users, where $\mathcal{O}$ represents the set of QPSK constellation. We assume that $\vect{s}\sim \mathcal{O}\mathcal{N}\left ( \vect{0}_M,\sigma_{s}^2\matt{I}_{M} \right )$. In this system 1-bit quantization at the transmitter $\mathcal{Q}_t$ as well as at the receiver $\mathcal{Q}_r$ is deployed. Therefore, in order to mitigate the IUI, we make use of a two-stage precoder consisting of the digital precoder $\matt P$ and the analog precoder $\matt D$. The use of the 1-bit quantizer at the transmitter $\mathcal{Q}_t$ delivers a signal $\vect{y}_{Q}$ that belongs to the set of $\{\pm 1 \pm \jim\}$. It means that the magnitude of the entry ${y}_{Q,n}$,  $n=1,...,N$, is constant and its phase belongs to $\{\frac{\pi}{4}, \frac{3\pi}{4}, \frac{5\pi}{4}, \frac{7\pi}{4}\}$. As a result, all the antennas end up getting the same power. To recover the information loss of the power allocation due to $\mathcal{Q}_t$, we employ an analog precoder $\matt D$ of real-valued diagonal structure. So, we end up with $\vect{y}_{Q_D} = \matt{D} \Q_t ( \matt{P} \vect{s} )$, where $\matt D \in \mathcal{R}^{N\times N}$, $\matt P \in \mathcal{C}^{N\times M}$ and $\matt P^{\T} =  \begin{bmatrix}\vect{p}_1 &\hdots &\vect{p}_N\end{bmatrix} $. The received decoded signal vector $\hat{\vect{s}} \in \mathcal C^{M \times 1}$ of the $M$ single-antenna users reads as $\hat{\vect{s}} = \Q_r (\matt H \vect{y}_{Q_D} + \boldsymbol{\eta})$, where $\matt H \in \mathcal{C}^{M\times N}$ is the channel matrix with i.i.d. complex-valued entries of zero mean and unit variance and $\vect{\boldsymbol{\eta}}\sim \mathcal{C}\mathcal{N}\left ( \vect{0}_M,\matt{C}_{\boldsymbol{\eta}} = \matt{I}_{M} \right )$ is the AWG noise vector.

\section{Statistical Theory of 1-bit Quantization}
\label{sec:onebitQ}

To design a precoder which takes into account the effects of $\mathcal{Q}_t$ and $\mathcal{Q}_r$, we need to know some statistical properties of quantization especially the auto and cross-correlation properties for a Gaussian input signal. Since quantization is non-linear, it has strong effects on the statistical properties of the signal. The statistical properties of hard limiters dealing with real-valued Gaussian-distributed signals are derived in \cite{Price1958}. These derivations are applied to complex-valued Gaussian-distributed signals and introduced in this section.
For a complex-valued signal vector $\textbf{x}$ we get
%\begin{subequations}
\begin{align}
&\Q(\textbf{x})=\Q\left ( \textbf{x}_{R}+ \jim \textbf{x}_{I} \right )=\Q \left ( \textbf{x}_{R} \right )+ \jim \Q\left (\textbf{x}_{I} \right ).
\end{align}
The covariance matrix between an unquantized circular distributed complex-valued signal $\textbf{x}$ of covariance matrix $\matt C_{\textbf{{x}}}$ and its 1-bit quantized signal $\textbf{x}_{\Q}=\Q \left ( \textbf{x} \right )$ is given by
%\begin{align}
%\E\left [ x_{1}\Q(x_{2})^{*} \right ]&= \sqrt{\frac{2}{\pi}}\left ( \sigma_{x_{{1}_R}}\rho_{x_{{1}_R}x_{{2}_R}}+\sigma_{x_{{1}_I}}\rho_{x_{{1}_I}x_{{2}_I}}\right. \nonumber\\
%&\left.-\jim\sigma_{x_{{1}_R}}\rho_{x_{{1}_R}x_{{2}_I}}+\jim \sigma_{x_{{1}_I}}\rho_{x_{{1}_I}x_{{2}_R}} \right ),
%\label{eq:Q1}
%\end{align}
\begin{align}
\textbf{{C}}_{\textbf{{x}}_{\Q}\textbf{{x}}}=\sqrt{\frac{4}{\pi}}\textbf{{K}}\textbf{{C}}_{\textbf{{x}}} \textrm{, where} \:\: \textbf{{K}}=\textrm{diag}\left ( \textbf{{C}}_{\textbf{{x}}} \right )^{-\frac{1}{2}}.
\label{eq:Q1}
\end{align}
The covariance matrix of the 1-bit quantized circular distributed complex-valued signals $\textbf{x}_{\Q}$ is given by
%\begin{align}
%\E\left [ \Q(x_{1})\Q(x_{2})^{*} \right ]&=\frac{2}{\pi}\left \{ \textrm{sin}^{-1}\left ( \rho_{x_{{1}_R}x_{{2}_R}} \right )+\sin^{-1}\left (\rho_{x_{{1}_I}x_{{2}_I}}  \right )\right. \nonumber\\
%&\left.-\jim \sin^{-1}\left (\rho_{x_{{1}_R}x_{{2}_I}}  \right )+\jim \sin^{-1}\left ( \rho_{x_{{1}_I}x_{{2}_R}} \right )\right \}.
%\label{eq:Q2}
%\end{align}
\begin{align}
\textbf{{C}}_{\textbf{x}_{\Q}}={\frac{4}{\pi}}\left ( \arcsin{\left ( \textbf{K}\Re\left \{ \textbf{{C}}_{\textbf{{x}}} \right \} \textbf{{K}}\right )}+\textrm{j}\arcsin{\left ( \textbf{{K}}\Im\left \{ \textbf{{C}}_{\textbf{{x}}} \right \} \textbf{{K}}\right )} \right ).
\label{eq:Q2}
\end{align}
Note that the diagonal entries of $\textbf{{C}}_{\textbf{{x}}_{\Q}}$ are the squared norm of quantized signals, which lead to the following result
\begin{align}
\textrm{diag}\left ( \textbf{{C}}_{\textbf{x}_{\Q}} \right )=2 \matt{I}.
\label{eq:Q3}
\end{align}
These four equations are the basis for solving the optimization problem presented in the next section.

\section{Optimization Problem}
\label{sec:optproblem}

The optimization problem is formulated as follows
\begin{align}
\begin{split}
\left \{ \matt{P}_{\text{MMSE-Q}},\matt{D}_{\text{MMSE-Q}} \right \}=\argmin_{\matt{P},\matt{D}} \: \E\left [ \left \|  \hat{\vect{s}}- \vect{s}\right \|_{2}^{2} \right ]\\
 \textrm{s.t.}\:\:\E\left [ \left \| \vect{y}_{Q_{D}} \right \|_{2}^{2} \right ]\leq E_{\text{tx}} \textrm{ and } \matt D \in \mathcal{R}^{N\times N} \textrm{is diagonal.}
\end{split}
\label{eq:optproblem}
\end{align}
\subsection{Objective function}
We aim at minimizing the MSE between the desired signals $\vect{s}$ and the received signals $\hat{\vect{s}}$ given that the power of the transmitted signal $\vect{y}_{Q_{D}}$ is limited by the available transmit power $E_{\text{tx}}$. We end up with the following expression for MSE 
\begin{align}
\text{MSE}\!=\!\sigma_s^2 \tr\left ( \matt{I}_{M}\! \right )\!+\!\tr\left (\E\left [\hat{\vect{s}}\hat{\vect{s}}^{H}\right ]\!-\! \E\left [\vect{s}\hat{\vect{s}}^{H}\right ]\!-\!\E\left [\hat{\vect{s}}\vect{s}^{H}\right ]\right).
\label{eq:mse}
\end{align}
We have to find the three expectation terms in the above for which we make use of the covariance and cross correlation matrices. We have already mentioned that the input signal covariance matrix $\matt{C}_{\vect{s}} = \sigma_{s}^2\matt{I}_M$. The covariance matrix of the precoder's output $\vect{y}$ is given by 
\begin{align}
\matt{C}_{\vect y}=\E\left \{ \vect{y}\vect{y}^{H} \right \}=\sigma_s^2\matt{P}\matt{P}^{H}.
\end{align}
To find a linear expression for the covariance matrix $ \matt{C}_{\vect{y}_{Q}}=\E\left \{ \vect{y}_{Q}\vect{y}^{H}_{Q} \right \}$, we make use of (\ref{eq:Q2}), (\ref{eq:Q3}) and the approximation $\arcsin{(x)}\simeq x, \textrm{ for } x \neq 1$. So, we get 
\begin{align}
%\matt{C}_{\vect{y}_{Q}}&=\begin{bmatrix} 2 &\cdots & \frac{4}{\pi}{\frac{\vect{p}_{1} \vect{p}^{H}_{N}}{\left \| \vect{p}_{1}  \right \|_{2}^{2}\left \| \vect{p}_{N}  \right \|_{2}^{2}}} \\
%\vdots & \ddots  & \\
%\frac{4}{\pi}{\frac{\vect{p}_{N} \vect{p}_{1}^{H}}{\left \| \vect{p}_{N}  \right \|_{2}^{2}\left \| \vect{p}_{1}   \right \|_{2}^{2}}}  & \cdots  & 2
% \end{bmatrix}  \label{eq:C_yq} \\
\matt{C}_{\vect{y}_{Q}}&=\frac{4}{\pi}\left [ \matt{K}_{2}  \matt{P}\matt{P}^{H} \matt{K}_{2} + c \matt I_{N}\right ],  \label{eq:C_yq}
\end{align}
%(\ref{eq:Q3}) is used to find the diagonal elements of $\matt{C}_{\vect{y}_{Q}}$. The non-diagonal entries are determined based on (\ref{eq:Q2}).
%The matrix $\matt{C}_{\vect{y}_{Q}}$ has entries that contain $\sin^{-1}$ terms which makes it difficult to find a closed form expression for it. Therefore, we make an approximation to remove the $\sin^{-1}$ terms. We approximate $\sin^{-1}{(x)}=x$ for a small value $x$. Using the mentioned approximation and mathematical derivations we obtain a closed form expression for $\matt{C}_{\vect{y}_{Q}}$
%\begin{align}
%\matt{C}_{\vect{y}_{Q}}=\frac{4}{\pi}\left [ \matt{K}_{2}  \matt{P}\matt{P}^{H} \matt{K}_{2} + c \matt I_{N}\right ], 
%\end{align}
with  $$\matt{K}_{2}=\diag\left ( \matt{P}\matt{P}^{H} \right ) ^{-\frac{1}{2}}$$ and $c =\left ( \frac{\pi}{2} -1\right )$.
The covariance matrix $\matt{C}_{\vect{y}_{Q_D}}$ of the transmitted signal $\vect{y}_{Q_{D}}$ is given by
\begin{align}
\matt{C}_{\vect{y}_{Q_D}}=\matt{D} \matt{C}_{\matt{y}_{Q}}\matt{D}.
\end{align}
The received signal covariance matrix reads as
\begin{align}
\matt{C}_\vect{x}=\matt{H}\matt{D}\matt{C}_{\vect{y}_{Q}}\matt{D}\matt{H}^{H}+\matt{C}_{\boldsymbol{\eta}}.
\end{align}
If we look at our MSE expression in (\ref{eq:mse}), one of the terms which we need to find is $\tr\left (\E\left [\hat{\vect{s}}\hat{\vect s}^{H}\right ] \right )$. Since the structure of $\matt C_{\hat{\vect s}}=\E \left [\hat{\vect s}\hat{\vect s}^{H}\right ]$ is very similar to $\matt C_{\textbf{\textit{y}}_{Q}}$, we end up with
\begin{align}
\tr\left (\E\left [\hat{\vect s}\hat{\vect s}^{H}\right ] \right )=2M
\label{eq:mse_T2}
\end{align}
We still need to find two more terms in the MSE, i.e. $\tr\left (\E\left [\hat{\vect s}\vect s^{H}\right ] \right)$ and $\tr \left ( \E\left [\vect s\hat{\vect s}^{H}\right ] \right)$ before we can proceed to solve for $\matt P$. Note that $\tr\left (\E\left [\hat{\vect s}\vect s^{H}\right ] \right)=\tr\left ( \E\left [\vect s\hat{\vect s}^{H}\right ] ^{H} \right )$. We use  (\ref{eq:Q1}) to calculate the above mentioned expectations, which can be expressed as follows
\begin{align}
\E\left [{\hat{\vect s}\vect s}^{H}\right ]&={\frac{4\sigma_s}{\pi}}\matt K_{1} \matt H\matt D\matt K_{2}\matt P 
\label{eq:mse_T3}\\
\E\left [{\vect s}\hat{\vect s}^{H}\right ] 
&={\frac{4\sigma_s}{\pi}}\matt P^{H} \matt K_{2}\matt D\matt H^{H}\matt K_{1},
\label{eq:mse_T4}
\end{align}
with $$\matt{K}_{1}=\diag\left ( \matt{C}_{\vect x} \right )^{-\frac{1}{2}}.$$ 
Finally, putting the expressions of (\ref{eq:mse_T2}), (\ref{eq:mse_T3}) and (\ref{eq:mse_T4}) in (\ref{eq:mse}), we end up with the following closed-form expression for MSE
\begin{align}
\textrm{MSE}&=\sigma_s^2M+2M \nonumber \\
&-{\frac{4\sigma_s}{\pi}}\tr\left (\matt K_{1} \matt H\matt D\matt K_{2}\matt P +\matt P^{H} \matt K_{2}\matt D\matt H^{H}\matt K_{1}   \right ).
\label{eq:MSE_1}
\end{align}
The MSE expression in (\ref{eq:MSE_1}) contains two unknown variables $\matt P$ and $\matt D$. Intuitively, $\matt D$ should be a function of $\matt P$, since it reallocates the power to the transmit signal originally intended by $\matt P$ which gets lost due to $\mathcal{Q}_t$. To this end, we define a new matrix $\matt P'=\matt K_{2}\matt P$. $\matt P'$ is a row-normalized version of $\matt P$, such that each row of $\matt P'$ has unit norm. Note that the MSE expression contains the product  $\matt K_{2}\matt P$ and $\matt P^{H}\matt K_{2}$. $\matt K_{1}$ also contains these products. Thus, the MSE expression found so far is in $\matt P'$ rather than $\matt P$. The purpose of  $\matt D$ is to remove this row-normalization of $\matt P$ introduced by $\matt K_{2}$. Therefore, an obvious choice for $\matt D$ is
\begin{align}
\matt D&=\matt K^{-1}_{2}=\textrm{diag}\left ( \matt P\matt P^{H} \right ) ^{\frac{1}{2}}.
\label{eq:D}
\end{align}
In other words, the optimization with respect to $\matt D$ is reformulated as a one with respect to the norm of each row of $\matt P$.
\subsection{Constraint}
The constraint can be simply expressed as
\begin{align*}
\E\left [\left \| \textbf{\textit{y}}_{Q_{D}}\right \|_{2}^{2} \right ]=\tr\left ( \matt D \matt C_{\textbf{\textit{y}}_{Q}}\matt D \right ) \leq E_{\text{tx}}.
\end{align*}
%Intuitively, in order to obtain the best performance we should use all the available transmit power i.e. $\E\left [\left \| \textbf{\textit{y}}_{Q_{D}}\right \|_{2}^{2} \right ] = E_{\text{tx}}$. 
Using (\ref{eq:C_yq}) and the fact that $\matt D $ is a diagonal matrix, we can simplify the constraint to 
\begin{align}
2\tr\left ( \matt D^2  \right ) \leq E_{\text{tx}}.
\label{eq:constraint}
\end{align}

\subsection{Final Optimization Problem}
Using (\ref{eq:MSE_1}), (\ref{eq:D}) and (\ref{eq:constraint}), we can finally write our optimization problem as
\begin{align}
\begin{split}
\min_{\matt P} \:\: &\sigma_s^2M+2M-{\frac{4\sigma_s}{\pi}}\tr\left (\matt K_{1} \matt H\matt P +\matt P^{H} \matt H^{H}\matt K_{1}   \right )\\
&\textrm{s.t.}\:\:\tr\left ( \matt P\matt P^{H} \right ) \leq \frac{E_{\text{tx}}}{2} \textrm{ and } \matt D = \textrm{diag}\left ( \matt P\matt P^{H} \right ) ^{\frac{1}{2}}.
\end{split}
\label{eq:final_opt}
\end{align}
%\begin{align}
%\begin{split}
%\matt{P}_{\text{MMSE-Q}}&=\argmin_{\matt{P}} \sigma_s^2M+2M-{\frac{4\sigma_s}{\pi}}\tr\left (\matt K_{1} \matt H\matt P +\matt P^{H} \matt H^{H}\matt K_{1}   \right )\\
%& \textrm{s.t.}\:\:\tr\left ( \matt P\matt P^{H} \right )=\frac{E_{\text{tx}}}{2}.
%\end{split}
%\end{align}
\subsection{Solving the Optimization Problem}
Note that our objective function in (\ref{eq:final_opt}) is non-linear in $\matt P$ because of $\matt K_1$. Furthermore, the solution set has to satisfy the constraint in (\ref{eq:final_opt}). Thus, we resort to the gradient projection algorithm to solve our optimization problem \cite{NLP}. The steps for this algorithm can be found in Table \ref{table}.  The needed derivative of the MSE with respect to $\matt P$ is given by
\begin{align*}
\begin{split}
\frac{\partial \textrm{MSE}\left (\textbf{\textit{P}}  \right )}{\partial \matt P}=-\frac{4}{\pi}\sigma_s &\left[\matt H^{\textrm{T}}\matt K_{1}-\frac{2}{\pi}  \matt H^{\textrm{T}}\matt K_{1}^3 \textrm{diag}\left ( \matt H^{*} \matt P^{*}\right ) \matt H^{*}\matt P^{*} \right.\vspace{0.5em}\\
&\left.-\frac{2 c}{\pi}  \textrm{diag}\left ( \matt H^{\textrm{T}}\textrm{diag}\left ( \matt H^{*} \matt P^{*}\matt K_{1}^3\right )\matt H^{*} \right )\matt P^{*} \right.\vspace{0.5em}\\
&\left. -\frac{2}{\pi}  \matt H^{\textrm{T}}\matt K_{1}^3 \textrm{diag}\left ( \matt P^{\textrm{T}} \matt H^{\textrm{T}}\right ) \matt H^{*}\matt P^{*} \right.\vspace{0.5em}\\
&-\left. \frac{2 c}{\pi}  \textrm{diag}\left ( \matt H^{\textrm{T}}\textrm{diag}\left ( \matt K_{1}^3\matt P^{\textrm{T}}\matt H^{\textrm{T}} \right )\matt H^{*} \right )\matt P^{*}\right].
\end{split}
\end{align*}
\begin{table}[]
\centering
\caption{Gradient Projection Algorithm}
\label{table}
\begin{tabular}{l}
\hline
Iteration step  $\mu$,Tolerable error $\epsilon$\\
\textbf{Initialization}$ \: \matt P_{(0)} = \matt H^{H}$, $n = 0$        \\
\textbf{Step 1:} If \:$\tr\left ( \matt P_{(0)} \matt P^{H}_{(0)}\right)> \frac{E_{\text{tx}}}{2}$, \\
\hspace{1.2cm} let $\matt P_{(0)}=\sqrt{\frac{0.5 E_{\text{tx}}}{\tr\left ( \matt P_{(0)} \matt P^{H}_{(0)} \right )}}\matt P_{(0)};$ \\
\textbf{Step 2:} $\matt P_{\left ( n+1 \right )}=\matt P_{\left ( n\right )}-\mu\left ( \frac{\partial \textrm{MSE}\left (\matt P_{(n)}  \right )}{\partial \matt P}\right )^{*}$;\\
\hspace{1.2cm} If \:$\tr\left ( \matt P_{(n+1)} \matt P^{H}_{(n+1)} \right ) > \frac{E_{\text{tx}}}{2}$, \\ 
\hspace{1.2cm} let $\matt P_{(n+1)}=\sqrt{\frac{0.5 E_{\text{tx}}}{\tr\left ( \matt P_{(n+1)}\matt P^{H}_{(n+1)} \right )}}\matt P_{(n+1)};$ \\
\textbf{Step 3:} If $\text{MSE}_{(n+1)} - \text{MSE}_{(n)} \leq \epsilon$ \\ \hspace{1.2cm} $\rightarrow$ terminate the algorithm.\\
\hspace{1.2cm} Otherwise, let $n=n+1$ and return to Step 2.\\
\hline
\end{tabular}
\end{table}
\section{Simulation Results}
\label{sec:simresults} 
In this section, we compare our proposed precoder with different precoding schemes in terms of the uncoded BER. All the simulation results are averaged over 200 channel realizations. The used modulation scheme is QPSK, where $\sigma^2_s =2$. $N=20$ antennas at the BS serve $M=4$ users with $N_b=1000$ transmit symbols per channel use. The tolerable error $\epsilon$ and the iteration step $\mu$ of the gradient projection algorithm are set to $10^{-6}$ and $0.05$, respectively. 

\begin{figure}[htb]
  \centering
  \centerline{\includegraphics[width=\columnwidth]{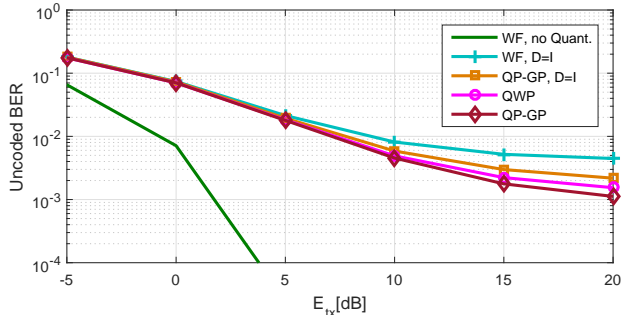}}
  \vspace{-0.5cm}
  \caption{\small{BER comparison between different precoding schemes}}
\label{fig:ber}
\end{figure}

In Fig. \ref{fig:ber} the uncoded BER is simulated as function of the available transmit power $E_{\text{tx}}$. "WF, no Quant." refers to the linear Wiener filter precoder while no quantization is applied in the system model. "WF, D=I" is the linear Wiener filter precoder that does not take the quantization into account and equal power allocation is performed \cite{Joham}. The transmit power constraint is still satisfied by appropriate scaling. "QP-GP" denotes our proposed precoder design: Quantized Precoder with Gradient Projection method. "QP-GP, D=I" refers to the proposed precoder design when the power allocation is equal for all transmit antenns. So, no additional analog processing $\matt D$ is required. "QWP" designates the Quantized Wiener filter Precoder introduced in \cite{Mezghani2009}. It can be seen from the results that ignoring the distortions due to the 1-bit quantization in WF leads to the worst case scenario. When taking them into account in QP-GP, D=I a significant improvement in the uncoded BER can be achieved. This performance improvement can be further increased when unequal power allocation at the transmit antennas is deployed, as shown in the case of QP-GP and QWF. The proposed precoder design QP-GP outperforms the other designs. This iterative design converges to the same solution for different initial values. 

\begin{figure}[htb]
  \centering
  \centerline{\includegraphics[width=\columnwidth]{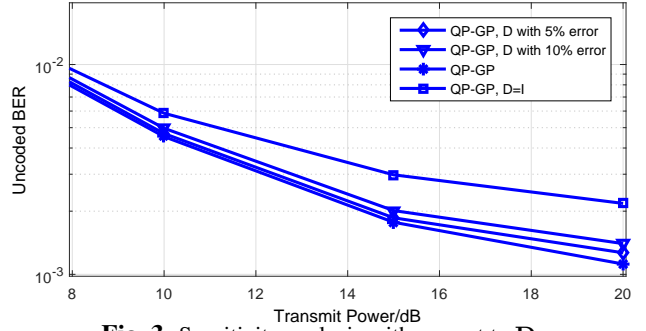}}
  \vspace{-0.5cm}
  \caption{\small{Sensitivity analysis with respect to $\matt D$}}
\label{fig:sensitivity}
\end{figure}

In general, the analog processing exhibits higher complexity and lower accuracy as compared to the digital counterpart due to hardware implementations and imperfections (aging, temperature,...). The analog precoder $\matt D$ offers less complexity due to its positive real-valued diagonal structure, and since it has to be updated only every coherence time. The BER performance sensitivity to inaccuracy in $\matt D$ implementation is studied and plotted in Fig. \ref{fig:sensitivity}. It can be seen that even with 10\% error in $\matt D$, the BER performance does not degrade much as compared to the ideal case. 

\begin{figure}[htb]
  \centering
  \centerline{\includegraphics[width=\columnwidth]{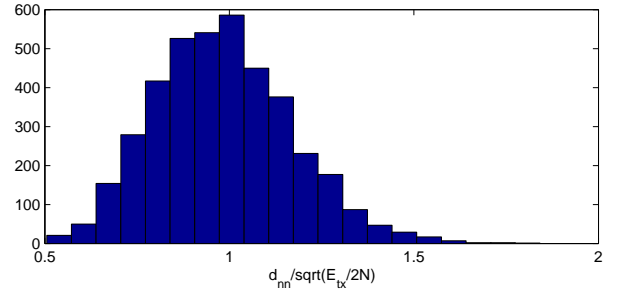}}
  \vspace{-0.5cm}
  \caption{\small{Distribution of the diagonal $\matt D$ coefficients over 200 channel realizations: $E_{\text{tx}}= 10$dB}}
\label{fig:D_coeff}
\end{figure}

The analog real-valued diagonal precoder $\matt D$ can be built within the power amplifiers at each antenna. Fig. \ref{fig:D_coeff} shows the distribution of the normalized diagonal coefficients of $\matt D$. We observe that the deviation of these coefficients among the different antennas and the channel realizations with respect to the mean value (at max 6dB) is quite small. Therefore, the requirements in terms of the dynamic range of the power amplifier are still reasonable.

\vspace{-0.5cm}
\section{Conclusion}
\label{sec:concl}
We present a new MMSE precoder design to mitigate the IUI in a DL massive MIMO scenario assuming perfect CSIT. The proposed precoder design takes into account the signal distortions due to the 1-bit quantization at the transmitter and at the receiver. The precoder is split into a digital precoder that separates the users in the direction and a real-valued diagonal analog precoder for the power allocation at each antenna. Our precoding  method shows better performance in terms of the uncoded BER compared to the precoder designed in \cite{Mezghani2009}.
The analog precoder involved in the proposed scheme is updated every coherence time thus reducing the implementation complexity. Furthermore, the BER performance is insensitive to imperfections in the analog precoder implementation.

\clearpage
\vfill\pagebreak
\bibliographystyle{IEEEbib}
\bibliography{refs}

\begin{thebibliography}{10}

\bibitem{Marzetta2010}
T.L. Marzetta,
\newblock ``Noncooperative cellular wireless with unlimited numbers of base
  station antennas,''
\newblock {\em Wireless Communications, IEEE Transactions on}, vol. 9, no. 11,
  pp. 3590--3600, November 2010.

\bibitem{Bjornson2013}
E.~Bjornson, M.~Kountouris, and M.~Debbah,
\newblock ``Massive mimo and small cells: Improving energy efficiency by
  optimal soft-cell coordination,''
\newblock in {\em Telecommunications (ICT), 2013 20th International Conference
  on}, May 2013, pp. 1--5.

\bibitem{Peel2005}
C.B. Peel, B.M. Hochwald, and A.L. Swindlehurst,
\newblock ``A vector-perturbation technique for near-capacity multiantenna
  multiuser communication-part i: channel inversion and regularization,''
\newblock {\em Communications, IEEE Transactions on}, vol. 53, no. 1, pp.
  195--202, Jan 2005.

\bibitem{Gershman2010}
A.B. Gershman, N.D. Sidiropoulos, S.~ShahbazPanahi, M.~Bengtsson, and
  B.~Ottersten,
\newblock ``Convex optimization-based beamforming,''
\newblock {\em Signal Processing Magazine, IEEE}, vol. 27, no. 3, pp. 62--75,
  May 2010.

\bibitem{Rusek2013}
F.~Rusek, D.~Persson, Buon~Kiong Lau, E.G. Larsson, T.L. Marzetta, O.~Edfors,
  and F.~Tufvesson,
\newblock ``Scaling up mimo: Opportunities and challenges with very large
  arrays,''
\newblock {\em Signal Processing Magazine, IEEE}, vol. 30, no. 1, pp. 40--60,
  Jan 2013.

\bibitem{Renzo2014}
M.~Di~Renzo, H.~Haas, A.~Ghrayeb, S.~Sugiura, and L.~Hanzo,
\newblock ``Spatial modulation for generalized mimo: Challenges, opportunities,
  and implementation,''
\newblock {\em Proceedings of the IEEE}, vol. 102, no. 1, pp. 56--103, Jan
  2014.

\bibitem{MullerSedaghat}
R.R. Muller, M.A. Sedaghat, and G.~Fischer,
\newblock ``Load modulated massive mimo,''
\newblock in {\em Signal and Information Processing (GlobalSIP), 2014 IEEE
  Global Conference on}, Dec 2014, pp. 622--626.

\bibitem{Kalis2012}
A.~Kalis, A.~Kanatas, and C.~Papadias,
\newblock {\em {P}arasitic {A}ntenna {A}rrays for {W}ireless MIMO {S}ystems},
\newblock Springer, 2014.

\bibitem{Bjornson2015}
E.~Bjornson, M.~Matthaiou, and M.~Debbah,
\newblock ``Massive {MIMO} with {N}on-{I}deal {A}rbitrary {A}rrays: Hardware
  scaling laws and circuit-aware design,''
\newblock {\em Wireless Communications, IEEE Transactions on}, vol. PP, no. 99,
  pp. 1--1, 2015.

\bibitem{Svensson2006}
C.~Svensson, S.~Andersson, and P.~Bogner,
\newblock ``On the power consumption of analog to digital converters,''
\newblock in {\em Norchip Conference, 2006. 24th}, Nov 2006, pp. 49--52.

\bibitem{Mezghani2009}
A.~Mezghani, R.~Ghiat, and J.A. Nossek,
\newblock ``Transmit processing with low resolution d/a-converters,''
\newblock in {\em Electronics, Circuits, and Systems, 2009. ICECS 2009. 16th
  IEEE International Conference on}, Dec 2009, pp. 683--686.

\bibitem{Price1958}
R.~Price,
\newblock ``A useful theorem for nonlinear devices having gaussian inputs,''
\newblock {\em Information Theory, IRE Transactions on}, vol. 4, no. 2, pp.
  69--72, June 1958.

\bibitem{NLP}
D.~P. Bertsekas and J.~N. Tsitsiklis,
\newblock {\em Parallel and Distributed Computation:Numerical Methods},
\newblock Prentice-Hall, 1989.

\bibitem{Joham}
M.~Joham,
\newblock {\em Optimization of Linear and Nonlinear Transmit Signal
  Processing},
\newblock Ph.D. thesis, Lehrstuhl f\"ur Netzwerktheorie und Signalverarbeitung,
  Technische Universit\"at M\"unchen, 2004.

\end{thebibliography}

\end{document}